\documentclass{aa}
\usepackage{astron,times,latexsym,graphics,graphicx}
\usepackage[dvips]{color}

\newcommand{\teff}{$T_{\rm eff}$}
\begin{document}

\title{M67-1194, an unusually Sun-like solar twin in M67
  \thanks{Based on data obtained at ESO-VLT, Paranal
Observatory, Chile, Program ID 082.D-0726(A)}.}

\author{Anna \"Onehag \inst{1}
\and    Andreas Korn  \inst{1}
\and    Bengt Gustafsson \inst{1}
\and    Eric Stempels   \inst{1}
\and    Don A. VandenBerg  \inst{2}}

\institute{Department of Physics and Astronomy,
           Uppsala Astonomical Observatory,
           Box 515, S-751\,20 Uppsala, Sweden
\and       Department of Physics and Astronomy, UVIC,
           Victoria, BC V8W 3P6, Canada}

\date{Submitted to A\&A}
\offprints{Anna \"Onehag,
\email{Anna.Onehag@fysast.uu.se}}

\authorrunning{\"Onehag et al.}
\titlerunning{M67-1194, an unusually Sun-like solar twin in M67}

\abstract
{
The rich open cluster M67 is known to have a chemical
 composition close to solar, and an age around 4\,Gyr. It
 thus offers the opportunity to check our understanding
of the physics and the evolution of solar-type stars in a cluster environment.
}
{
We present the first spectroscopic study at high resolution, $R \approx $
50,000, of the potentially best solar twin, M67-1194, identified among
solar-like stars in M67.
}
{
G dwarfs in M67 ($d \approx$ 900 pc) are relatively
faint ($V \approx$ 15), which makes detailed spectroscopic studies
time-consuming.
Based on a pre-selection of solar-twin candidates performed at medium
resolution  by Pasquini et al. (2008), we explore the 
chemical-abundance similarities and differences between  M67-1194 and the Sun,
using VLT/FLAMES-UVES.

Working with a solar twin in the framework of a differential analysis, we
minimize systematic model errors in the abundance analysis compared to
previous studies which utilized more evolved stars to determine the
metallicity of M67. The differential approach yields precise and accurate
chemical abundances for M67, which enhances the possibility to
use this object in studies of the potential peculiarity, or normality, of the
Sun.
}
{
 We find M67-1194 to have stellar parameters indistinguishable from the
solar values, with the exception of the overall metallicity which is
slightly super-solar ([Fe/H] = 0.023 $\pm$ 0.015). An age determination
based on evolutionary tracks yields 4.2 $\pm$ 1.6\,Gyr. Most surprisingly,
we find the chemical abundance pattern to closely resemble the solar one,
in contrast to most known solar twins in the solar neighbourhood.}
{We confirm the solar-twin nature of M67-1194, the first solar twin known
to belong to a stellar association. This fact allows us to put some constraints
on the physical reasons for the seemingly systematic departure of M67-1194 and
the Sun from most known solar twins regarding chemical composition. We find
that radiative dust cleansing by nearby luminous stars may be the
explanation for the peculiar composition of both the Sun and M67-1194,
but alternative explanations are also possible.
The chemical similarity between the Sun and M67-1194 also suggests that
the Sun once formed in a cluster like M67.}


\maketitle
\section{Introduction}
\label{sec:intro}
For at least four decades,
searches have been conducted for stars with properties very similar to the Sun
(see Hardorp 1978, Cayrel de Strobel et al. 1981; see also the review by
Cayrel de Strobel 1996,
and for a recent short summary e.g. Mel{\'e}ndez \& Ram{\'i}rez 2007).
It would be important to find a star with physical characteristics
indistinguishable from those of the Sun, a ``perfect good solar twin'',
as defined by Cayrel de Strobel (1996). The reasons for this
importance are both physical and technical. Physically,  the
statistics of solar twins would
certainly contribute to our understanding of the uniqueness or
normality of the Sun (cf.
Gustafsson 1998). Technically, a solar twin would be useful in
setting zero points in the calibration of effective-temperature scales, 
based on stellar
colours. Another use
would be in the calibration of night-time reflectance spectroscopy of
solar-system bodies, where the
spectral component of the Sun must be removed before an analysis of
the spectroscopic
features of the body itself can be performed.

Numerous searches and accurate analyses have resulted in a small
sample of
solar-twin candidates (Porto de Mello \& da Silva 1997, Mel{\'e}ndez et
al. 2006, Takeda et al. 2007, Mel{\'e}ndez et al. 2009,
Ram{\'i}rez et al. 2009). Although these stars in general have
fundamental parameters very close to solar, recent advances in
high-accuracy differential abundance analyses have proven almost all
of them to have chemical
compositions slightly, but systematically, deviating from that of the Sun.

A special opportunity in the search for solar twins is offered by the
old and rich open cluster M67.
It has a chemical composition similar to the Sun with
[Fe/H] in the range $-$0.04 to 0.03 on the customary logarithmic
scale normalised to the Sun (Hobbs \& Thorburn 1991,
Tautvai$\check{\rm s}$iene et al.
2000, Yong et al. 2005, Randich et al. 2006, Pace et al. 2008,
Pasquini et al.
2008). Its age is also comparable to that of the Sun: 3.5\,--\,4.8 Gyr
(e.g.\ Yadav et al. 2008).
M67 is  relatively nearby ($\sim$\,900\,pc) and is only little
affected by interstellar extinction,
which allows detailed spectroscopic studies of its main-sequence
stars. The depleted Li abundance of the Sun seems rather representative
of solar-twins in the Galaxtic field (Baumann et al. 2010). M67 also seems
to contain Li-depleted G stars (Pasquini et al. 1997). 
M67 thus offers good possibilities of finding solar-twin
candidates for further exploration.
Pasquini et al. (2008) (followed by a paper of Biazzo et al.
2009) recently searched the cluster for solar analogs, and listed ten
promising candidates.

\begin{figure*}
  \centering
   \includegraphics[angle=90,width=18.5cm]{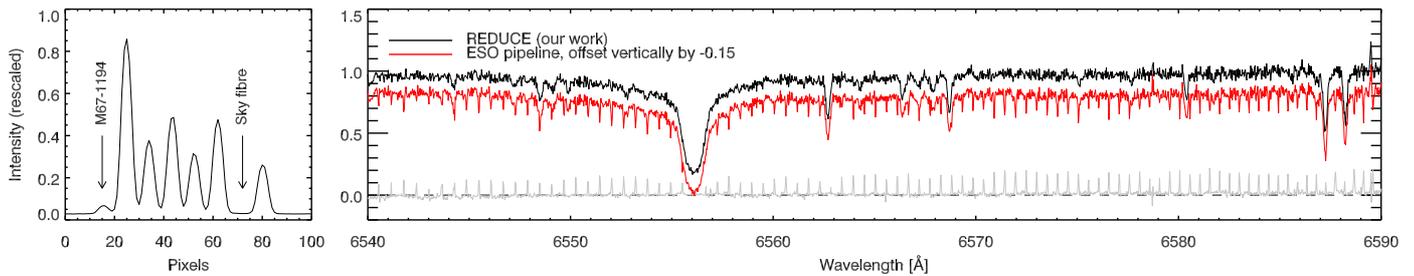}
  \caption{Left panel: Cross-dispersion cut showing the eight-fibre
    set-up. Right
  panel: The pipeline-reduced data
  provided by ESO (red) compared to  our work. Below the two spectra
  a difference spectrum is displayed for a direct comparison of the
  two reductions (data taken 2009-04-03: OB ID 331412).}
  \label{fig:reduction}
\end{figure*}

Here we present an analysis of  \textit{M67-1194}
(NGC 2682 YBP 1194, ES 4063, ES IV-63, FBC 2867, MMJ 5357, SAND 770, 
2MASS J08510080+1148527), a cluster
solar-twin candidate suggested by Pasquini et al. (2008). Our analysis 
is based on high-resolution
observations with relatively high signal-to-noise (S/N) ratio.
In Section \ref{sec:obs}, we discuss the
observations and some aspects of the data reduction.
Section \ref{sec:analys} describes
the analysis method and the determination of fundamental parameters
(\teff, $\log g$, [Fe/H] and $\xi_{t}$). In Section \ref{sec:comp} we present
the results of a detailed analysis of a
number of chemical elements, and these results are compared to those obtained
for known twins in the Galactic field. 
In Section \ref{sec:age}, we present a new age determination for M67,
and in Section \ref{sec:disc} we discuss the results.

\section{Observations}
\label{sec:obs}
The observations of M67-1194 were carried out with the multi-object
spectrometer FLAMES-UVES at ESO-VLT in Service Mode
in the spring of 2009 during
a period of three months (18th of January -- 3rd of April). The observations
analysed here are part of a larger project with a main goal to
study atomic diffusion in stars of M67
(082.D-0726(A)). In each observing block of the project,
one fibre of the spectrograph system was positioned on M67-1194 in order to
collect as many observations as possible of this faint G dwarf.

We obtained altogether 18\,h in 13 observing nights (23 individual
observations).
The spectrometer setting (RED580) was chosen to yield a resolution
of $R = \lambda/\Delta\lambda =$\,47,000 ($1''$ fibre) and a wavelength
coverage of 480--670\,nm.
The typical signal-to-noise ratio (S/N) per frame is
36\,pixel$^{-1}$ (as measured in the line-free region between 
6426\,\AA\ and 6430\,\AA).
Barycentric radial velocities were determined from the individual spectra
yielding a radial velocity of $-33.8 \pm 0.4$\,km/s.
This is in exellent agreement (1$\sigma$)
with the mean radial velocity of the cluster as
determined by Pasquini et al. (2008), $v_{\rm rad}=-33.30 \pm 0.40$,
and Yadav et al.\ (2008), $v_{\rm rad}=-33.673 \pm 0.208$.
The latter study classifies M67-1194 as a
high-probability proper-motion cluster member (99\%).
The frames were subsequently co-added for highest possible S/N ratio.

\subsection{Data reduction}
While ESO provides pipeline-reduced data for an initial assessment of the
observed spectra, these are not intended for a full scientific analysis. In
fact, we found that the extraction of our spectra by the ESO pipeline was
problematic, mainly because the instrumental setup of FLAMES-UVES combined with
our choice of targets is challenging. Having eight fibres tightly packed in a
slit-like fashion, FLAMES-UVES can obtain echelle spectra of up to 8
targets in parallel. However, the spacing between the individual fibres is
rather small, which leads to non-negligible light contamination between
adjacent fibres. This was made even more challenging by the fact that our
target is considerably fainter than the one of the adjacent fibre. The ESO
pipeline had severe problems extracting such a spectrum, which is apparent
from a strong and periodic pattern of low-sensitivity spikes. This is
illustrated in Fig.~\ref{fig:reduction}, where we show a cross-order
profile and a section of the pipeline-reduced spectrum of M67-1194 (in red) 
from frame
OB\,331412 (2009-04-03) --- an illustrative frame in this context.

Given the problems in the pipeline-reduced data, we adapted the code of the
echelle-reduction package REDUCE (Piskunov \& Valenti 2002) to
perform an accurate spectrum extraction.
Here we used special calibration frames provided by ESO, the so-called
{\it even} and {\it odd} flat-field frames. In these frames, only every second
fibre is illuminated, which eliminates crosstalk between fibres and allows us
to calculate an accurate two-dimensional model of the individual aperture
shapes.
The extraction of the target spectra was then performed in two steps. We first
used a very narrow aperture to extract the central (non-overlapping) part of
each fibre. This allowed us to estimate the relative contribution of each
individual fibre at each wavelength, which we then combined with the earlier
determined aperture shapes to create a full two-dimensional model (which
includes
spectral features) for each fibre in the observed frame. We then eliminated the
fibre-to-fibre contamination by subtracting the models of adjacent apertures
prior to a full extraction of each individual spectrum. We show an example of a
spectrum extracted with this method in  Fig.~\ref{fig:reduction} (black). The 
periodic
pattern of spikes has been removed successfully, and the agreement with the
pipeline-reduced spectra outside the spikes is excellent. 

\begin{figure*}
  \hspace{-0.4cm}
  \centering
   \includegraphics[angle=90,width=18.0cm]{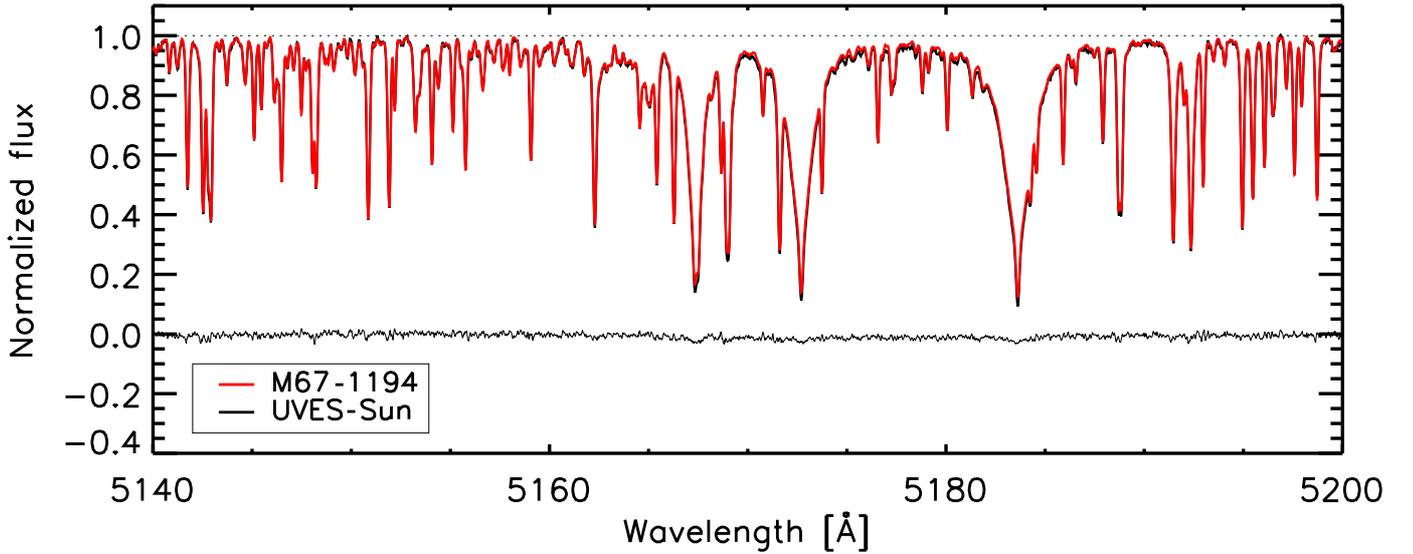}
  \caption{Observations of the Mg\,{\sc i}$b$ triplet region, for both
    M67-1194 (black) and the FLAMES-UVES Sun (red). A difference spectrum is 
    plotted below.}
  \label{fig:mgtriplet_global}
\end{figure*}

Fig.~\ref{fig:mgtriplet_global} shows a portion of the co-added spectrum 
covering the Mg\,{\sc i}$b$ lines. The S/N in the co-added spectrum reaches a 
mean of 160\,pixel$^{-1}$.

\section{Analysis}
\label{sec:analys}
We performed a line-by-line differential analysis of the combined spectrum of
M67-1194 relative to the Sun. The latter was represented by a day-time
spectrum of the sky taken with the 580\,nm setting of FLAMES-UVES in 2004
(Program ID 60.A-9022). The S/N of the solar spectrum is generally higher than
that of M67-1194 (typically 230\,pixel$^{-1}$). Together with the Kitt-Peak Solar Atlas
(Kurucz et al. 1984), this made it
possible to identify blends of weak lines, in the wings of the lines
measured as well as their surrounding regions used for setting the continuum.

\begin{figure}
  \hspace{-0.4cm}
  \centering
   \includegraphics[angle=90,width=9.3cm]{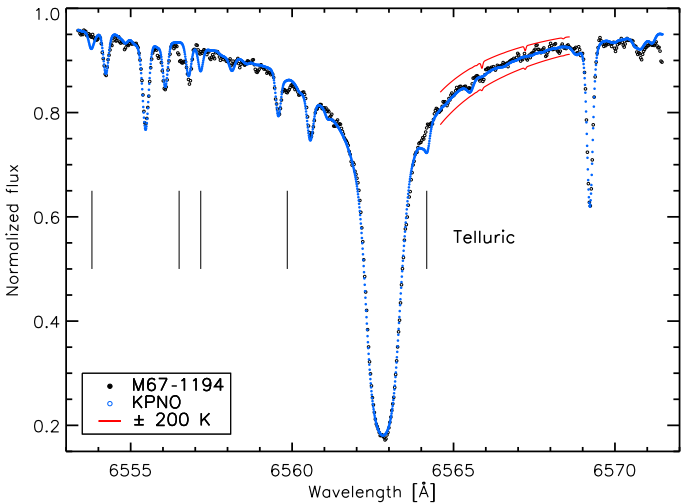}
  \caption{The H$\alpha$ line observed in the solar spectrum
    (Kitt Peak National Observatory, KPNO) and in the spectrum of M67-1194.
    Synthetic spectra with effective temperatures offset by $\pm$\,200\,K
    are shown as solid lines. The most prominent telluric lines are
    marked out as vertical lines.}
  \label{fig:halpha}
\end{figure}

\begin{figure}
  \hspace{-0.4cm}
  \centering
   \includegraphics[angle=90,width=9.3cm]{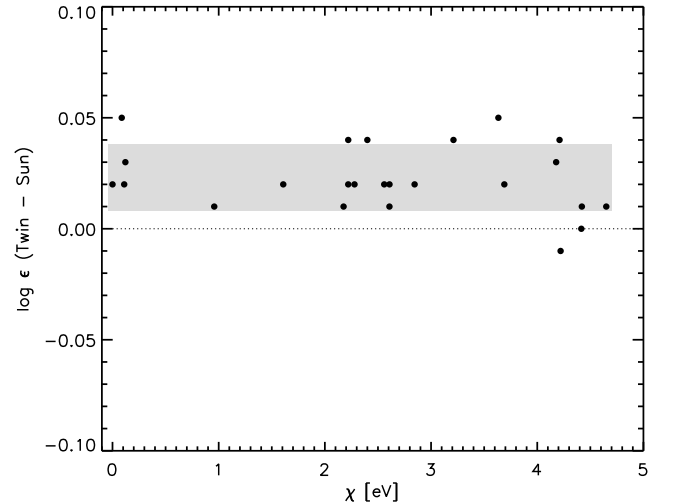}
  \caption{The line-by-line difference in Fe\,{\sc i} abundance
    plotted as a function
    of the excitation energy of each line. A slope insignificantly different
    from zero is found, suggesting an effective temperature close to 5780\,K.
    The standard deviation is indicated by the grey-shaded area.}
  \label{fig:eex}
\end{figure}

For analysis we used SIU (Reetz 1991), a tool to visualize and compare
observed and theoretical spectra. SIU has a built-in line-synthesis module
which uses one-dimensional hydrostatic model atmospheres in LTE with an ODF
representation of line opacity (MAFAGS, Fuhrmann et al. 1997, Grupp 2004). The
line-formation code also assumes LTE. The highly differential character of the
analysis make our results very little dependent on the
model atmospheres used; if, e.g., the more detailed MARCS models (Gustafsson
et al. 2008) were used, the same results would be obtained.

In a first step, we analysed the solar spectrum. We selected iron lines
from the work of Korn et al. (2003), complemented by lines used by
Mel\'endez et al. (2009, priv. comm.) to cover a wide range in
excitation energies. The KPNO solar atlas was consulted when setting the local 
continuum around lines of interest.
With \teff = 5777 K, log g = 4.44, [Fe/H] = 0.00,
a solar microturbulence of $\xi_t$ = 0.95 km/s was found to minimize trends of 
abundance with line strength for both Fe\,{\sc i} and Fe\,{\sc ii}.

The derivation of a differential elemental abundance is a three-step process.
First, the spectral region around every spectral line of interest is
carefully normalized by comparing the relative fluxes in the solar and the
stellar spectrum. Next, the solar spectrum is analysed varying the
product of the elemental abundance and the $gf$ value until a satisfactory fit 
to the line profile is achieved. For strong lines damping parameters are
based on quantum-mechanical calculations and adopted from VALD 
(Kupka et al. 1999). The Mg~{\sc i}$b$ triplet line, used in the determination 
of
the surface gravity, is fitted to the solar line varying the damping.
The treatment of H$\alpha$ follows Korn et al. (2003).
The ``external'' profile, including effects of macroturbulence, projected 
rotational velocity and instrumental profile, is assumed to be Gaussian with 
a width determined in this fitting process. Finally, the stellar spectrum is 
analysed and the abundance difference noted. Steps 2 and 3 are not carried out 
one after the other, to
avoid biases that interactive differential line fitting may be subject to.
Rather, a set of lines is analysed in the solar spectrum, then the same set
of lines in the stellar one, without preconceptions as regards the 
abundance.

We performed test calculations to quantify to which extent this 
interactive procedure gives results in agreement with those of a fully 
automatic procedure. An unweighted 
$\chi^2$ on the whole line profile was found to give systematically somewhat 
higher abundances in the analysis of the stellar spectrum. 
This is not surprising, as the experienced spectroscopist will tend to
compensate for suspected blends while $\chi^2$ will blindly fit them
as part of the line profile.
However, the offset in abundance never exceeded 0.01\,dex. Given the good 
agreement between these two fitting techniques, we are confident that the 
results of this study are independent of the line-fitting method used.

\begin{figure}
  \hspace{-0.4cm}
  \centering
   \includegraphics[angle=90,width=9.3cm]{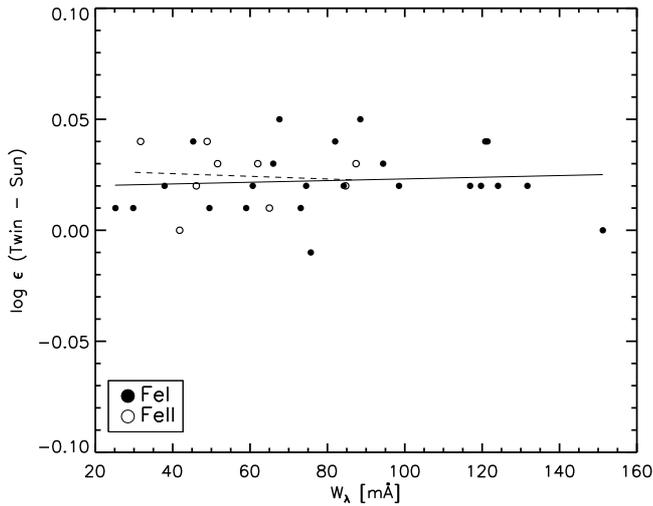}
  \caption{The line-by-line difference in Fe\,{\sc i} and
      Fe\,{\sc ii} abundances plotted as
    a function of the equivalent width of the lines. Fe\,{\sc i} lines return a
    slightly positive dependence (solid line), Fe\,{\sc ii} (dashed line) a 
    slightly negative one suggesting an average $\xi_t$ very close to the 
    solar value of 0.95\,km/s.}
  \label{fig:weq}
\end{figure}

\begin{figure}
  \hspace{-0.4cm}
  \centering
   \includegraphics[angle=90,width=9.3cm]{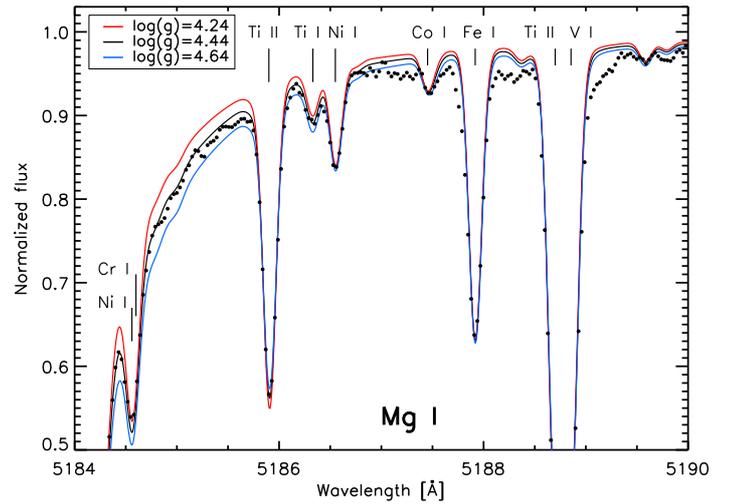}
  \caption{Observations of the gravity-sensitive wing of Mg\,{\sc i} 
    5183.6\,\AA\ shown together with synthetic spectra for models with 
    $\log g$ values of 4.24, 4.44, and 4.64 (cgs units).}
  \label{fig:mgtriplet}
\end{figure}

\subsection{Fundamental stellar parameters and chemical composition}
The stellar parameters of M67-1194 have been constrained employing several
spectroscopic techniques. The primary method to estimate the 
{\em effective temperature}
is based on the self-broadened wings of
H$\alpha$. As Fig.~\ref{fig:halpha} shows, the agreement between the H$\alpha$
profile observed for M67-1194 and that of the Kitt Peak atlas is excellent. 
Within the noise level, there
are no appreciable departures, except for the few telluric lines present in our
observations. The estimated
effective temperature of M67-1194 when based on the wings of H$\alpha$ is 
5780\,$\pm$\,50\,K.

We furthermore analysed the excitation equilibrium of neutral iron 
(see Fig.\,\ref{fig:eex}) and obtained a $T_{\rm eff}$ equal to 5780,
omitting one uncertain high-excitation line. 
We derived a mean metallicity for M67-1194 of 
[Fe/H]$_{\rm I}$ = 0.023 $\pm$ 0.015 
(1$\sigma$). The uncertainty in [Fe/H] translates into an uncertainty in 
effective temperature at the 20\,K level (1$\sigma$). 

Combining the above two indicators, we adopt 5780 $\pm$ 27\,K as our 
current best estimate of the effective
temperature of M67-1194. The error is the combined error from the two 
methods propagated into the mean.

Since the effective temperature is of key significance for the abundance 
results to be discussed below, we have performed two independent checks. One 
is to apply the line-ratio method of Gray (1994). Here, the central line depths
of two neighbouring spectral lines, one at low and one at relatively high 
excitation, are compared. Following Gray, we have chosen the V\,{\sc i} 
6251.8\,\AA\ line ($\chi$ = 0.29\,eV) and the Fe\,{\sc i} 6252.6\,\AA\ line 
($\chi$ = 2.40\,eV) 
which, assuming a solar [V/Fe] (consistent with the results of our analysis), 
suggests a temperature difference 
$\Delta = T_{\rm eff}$(M67-1194)$ - T_{\rm eff}(\odot) = 60 \pm 70\,$K.

A second method is to use the observed colour of the star. According to Yadav 
et al. (2008), $(B-V)_{\rm 1194} = 0.667 \pm 0.01$. Together with the 
comprehensive assessment of the reddening towards M67 
$(E(B-V) = 0.041\pm 0.004$, Taylor 2007), this points towards a $(B-V)$ of 
$0.626 \pm 0.011$ which is in the lower end of the interval
$(B-V)_{\odot}=0.62-0.65$ suggested by contemporary studies (see Holmberg et al.
2006, who, however,
favour a high value around 0.64). If we adopt the recent calibration of 
$(B-V)(T_{\rm eff})$ by Casagrande et al. (2010) we get 
$T_{\rm eff}$(M67-1194) = 5832\,K. If we adopt
a stellar $K$ magnitude from the 2MASS catalogue we find $(V-K)$ = 1.664, 
correct for reddening following Cox (2000) which  
suggests $(V-K)_0$ = 1.556, and apply the Casagrande et al. $(V-K)$
calibration we obtain $T_{\rm eff}$(M67-1194) = 5773\,K. 
Taking uncertainties in continuum 
positioning for the line-depth estimates and in reddening for the colour 
estimates into consideration, all these results are consistent with our 
adopted value of 5780\,K. The line-depth ratio method and photometry suggest 
that this value might rather be slightly underestimated than overestimated.

The {\em surface gravity} is set by requiring lines of Fe\,{\sc ii} to return 
the same differential abundance as lines of Fe\,{\sc i} (LTE ionization 
equilibrium).
As in the case of Fe\,{\sc i}, we opted for a careful selection of relatively 
few well-observed weakly blended lines.
Fig.~\ref{fig:weq} demonstrates the ionization equilibrium is established
at log $g$ = 4.44. Given a 1$\sigma$ scatter among the log $g$-sensitive 
Fe\,{\sc ii} lines (0.013\,dex), the value of the surface gravity is uncertain 
at the 0.04\,dex level. In this 
step, the microturbulence is found to be $\xi_t$ = 0.95 km/s (cf. 
Fig.~\ref{fig:weq}), i.e. the same value as used in the analysis of the solar 
spectrum.

Further evidence as regards the solar-like surface gravity is obtained from the analysis of the damping wings
of the Mg\,{\sc i}$b$ lines.
The Mg abundance was derived from the Mg\,{\sc i} doublet lines at 6718.7 and 6319.2\,\AA\ which are on the linear part
of the curve of growth, as well as from the moderately strong Mg\,{\sc I} 
5711\,\AA\
line. These lines consistently give the result [Mg/Fe] = 0.02 $\pm$ 0.01. There
are no other suitable atomic Mg lines in the spectral regions observed here.
To get further constraints on the Mg abundance, we explored the wavelength
region around 5134--5137\,\AA\ with Q$_{2}$(23), Q$_{1}$(23) and R$_{2}$(11)
features of the (0,0) bands of the A-X system of MgH. This region is
severely affected also by lines from the C$_{2}$ Swan bands. The observed
spectrum appears very similar but with about 8\% stronger MgH bands, as
compared with the solar spectrum.
Our spectrum synthesis, assuming solar isotopic ratios, shows that the observed
MgH bands are consistent
with our abundance estimate for Mg and the somewhat enhanced overall 
metallicity. The adoption of a significantly lower
Mg abundance (by 0.1 dex) would, however, lead to too weak MgH
features. Adopting [Mg/H] = 0.02, the pressure-broadened red wing of
Mg\,{\sc i} 5183\,\AA\ (the strongest line of the Mg\,{\sc i} triplet with the largest
log $g$ sensitivity)
indicates a log $g$ of 4.44 $\pm$ 0.05 (see Fig.~\ref{fig:mgtriplet}). To this
we add in quadrature the uncertainty stemming from the Mg abundance. This yields an error for $\log g$ of 0.06.
We finally adopt 4.44 $\pm$ 0.035 as our current best estimate of the surface
gravity of M67-1194. The error is the combined error from the two methods,
Fe\,{\sc i/ii} and Mg\,{\sc i}$b$, have been propagated into the mean.

\textit{Abundances} are determined using the line list of
Mel{\'e}ndez et al. (2009), which was chosen in order to make our results
directly comparable to the twin results of that group. In the case of carbon,
we needed to depart from that list replacing lines not available in our
wavelength range with C$_2$ lines.
In the case of oxygen, inspection of the wavelength regions around the
[O{\sc i}] lines at 6300.3 and 6363.7\,\AA\, proved this region to be far
too noisy for an accurate oxygen abundance determination. Within this 
noise, the spectrum does not depart significantly from the solar spectrum, 
and fits
of synthetic spectra suggest the oxygen abundance ratio [O/Fe] to be
solar with an estimated maximum error of about 0.15\,dex. The
high excitation O\,{\sc i} 6158.1 line was found
to be more useful at this level of S/N, allowing an estimated 
[O/Fe] = 0.07 $\pm$ 0.07.
These errors are here to be seen as maximum
errors, and the uncertainties are mainly related to the placement of the
continuum. We note that this result is consistent with the
mean [O/Fe] = 0.01 $\pm$ 0.03, as determined for ten somewhat brighter and 
hotter main-sequence stars in M67 by Randich et al. (2006) using the 
6300.3\,\AA\ line.

\begin{figure}
  \centering
   \includegraphics[width=9.3cm]{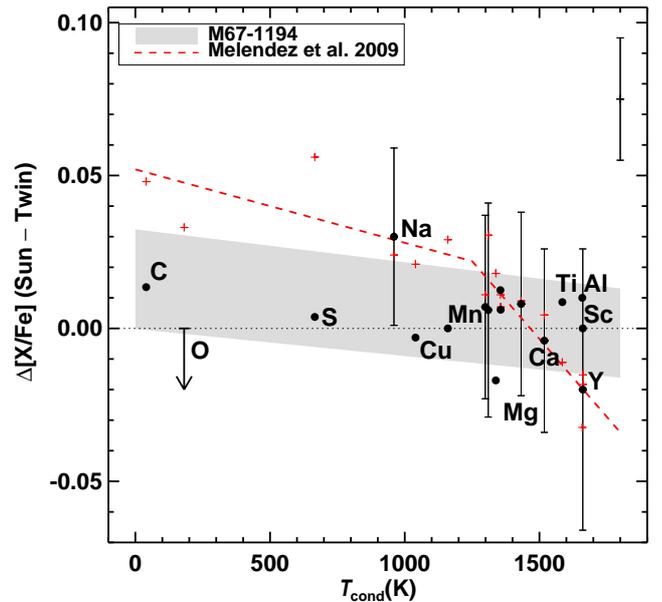}
  \caption{Abundance as a function of condensation temperature
    (according to Lodders 2003). Filled circles: this paper, plus signs:
    Mel{\'e}ndez et al. 2009. Shaded area: this paper, dotted line: Mel\'endez
    et al. 2009. The shaded area summarizes the result of this paper when 
    submitted to a linear-regression analysis, giving 
    $\Delta{\rm [X/Fe]} = a + m T_{cond}$. The area delineates the range of 
    regression lines due to the uncertainties in $a$ and $m$. The elements in 
    the range 1300--1500\,K not identified are (from left to right): Cr, Si, 
    Co, Ni and V. The error bars indicate the line-to-line scatter in 
    abundances, e.g. the internal errors. If this error is below 0.02, we adopt
    0.02\,dex (indicated in the upper right corner and representative
    of all points without error bars). In the case of Na (two lines giving
    abundances in numerical 
    agreement) we adopted an estimate of the fitting error.  The effects of 
    errors in fundamental stellar parameters lead to vertical shifts of the 
    points by typically 0.00\,--\,0.02\,dex.}
  \label{fig:diff}
\end{figure}

The resulting abundances are illustrated in Fig.~\ref{fig:diff} as differences 
between the solar values and those of M67-1194, and compared to the mean 
results for the solar twin sample of Mel{\'e}ndez et al. (2009).
We note that, given the similarity of the spectrum of M67-1194 to the
solar one, uncertainties in line data effectively cancel. Similarly, errors 
due to inadequacies in the model atmospheres, e.g., in describing the formation
of molecular lines in the upper layers of the atmospheres, should cancel,
provided that the elemental abundances are practically the same, and no
other effects but those related to the stellar fundamental parameters
characterize the atmospheres.

\begin{figure*}
  \hspace{-0.6cm}
  \centering
   \includegraphics[width=8.0cm,angle=90]{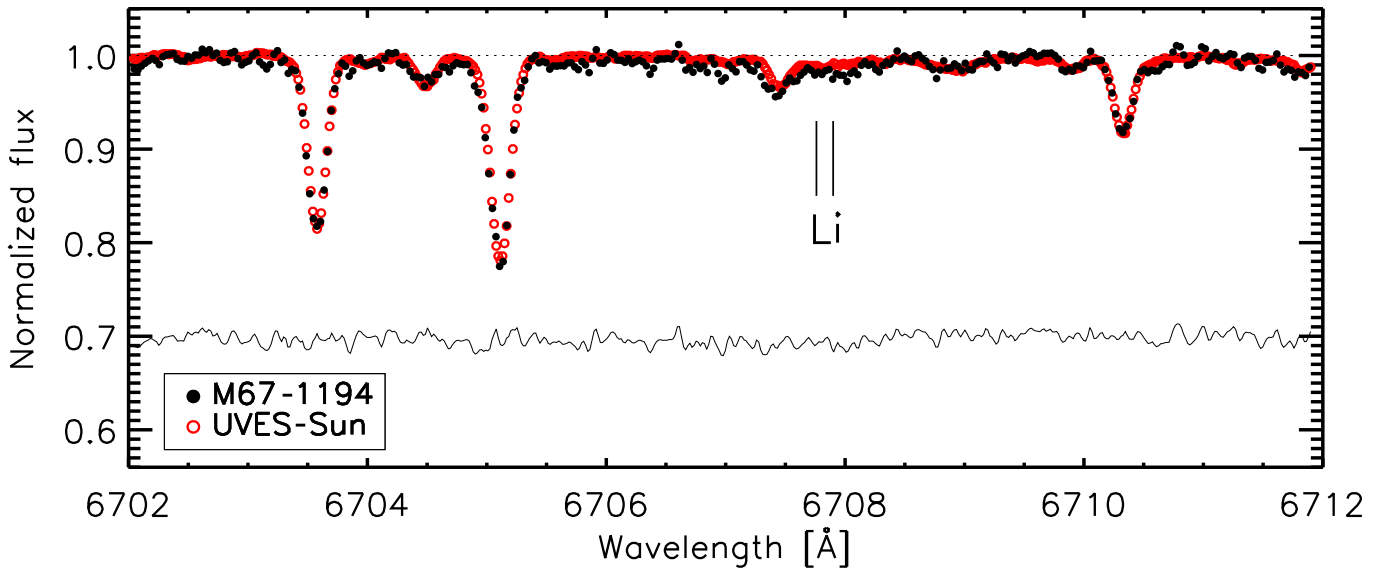}
  \caption{M67-1194 compared to the FLAMES-UVES spectrum
    representing the Sun (red). The difference spectrum is shown
    below (offset by +0.7 units). At the current S/N level, there is no 
    systematic difference between the two spectra at the position of the 
    Li\,{\sc i} doublet.}
  \label{fig:li_glob}
\end{figure*}

\section{M67-1194 in comparison with other twins}
\label{sec:comp}
The resulting abundance ratios for M67-1194 relative to iron and
normalized on the solar abundances,
are plotted in Fig.~\ref{fig:diff} versus the 50\% condensation temperature
of the various elements, according to Lodders (2003). Here, the
results of Mel\'endez et al. (2009) for their sample
of eleven solar twins in the Galactic field are also indicated, and following 
those authors we plot the differences between the solar abundances and 
corresponding results for the stars. 

Mel\'endez et al.\ found that for almost all their stars, the volatile 
elements are relatively less abundant than in the Sun, while the
refractory elements are somewhat enhanced in the stars. This effect
is seemingly {\it not} present in the M67 twin, at least not to the same
extent. It is thus seen (c.f. 
Fig.~\ref{fig:diff}) that the abundance profile of
M67-1194 is more similar to the Sun than for all the Mel\'endez et al.\ twins,
with one or two exceptions.   
We perform a simple statistical test, essentially judging the probability 
that all black dots in Fig.~\ref{fig:diff} just by chance fall above the red 
dashed line for
$T_{\rm cond}>$\,1400\,K, and below it for $T_{\rm cond}<$\,1400\,K. We find that
the probability for M67-1194 to be drawn from the same sample of stars as the
majority of the Mel\'endez et al.\ twins is less than 5\%. On the other hand, 
the errors in the analysis are fully compatible with solar abundance ratios for
the M67 star. This astonishing result will be discussed further below.

One could ask to which extent the result shown in Fig.~\ref{fig:diff} 
would change if the fundamental parameters of M67-1194 are in error.
We have found that the only error which could qualitatively affect the  
result would be an overestimate of the stellar temperature
by 50\,K or more. If the star is cooler by that amount, the  
abundances of the volatiles are overestimated so that when correcting
for that error the points for C, O and S in the diagram would be  
pushed upwards towards the red dashed "Mel{\'e}ndez twin" line while some  
points representing refractories in the right part of the diagram  
would be pushed down. In view of the discussion of possible
errors in $T_{\rm eff}$(M67-1194) we find, however, such an overestimate  
improbable.\\

\subsection{Lithium content}
\begin{figure}
  \hspace{-0.6cm}
  \centering
   \includegraphics[width=7.0cm,angle=90]{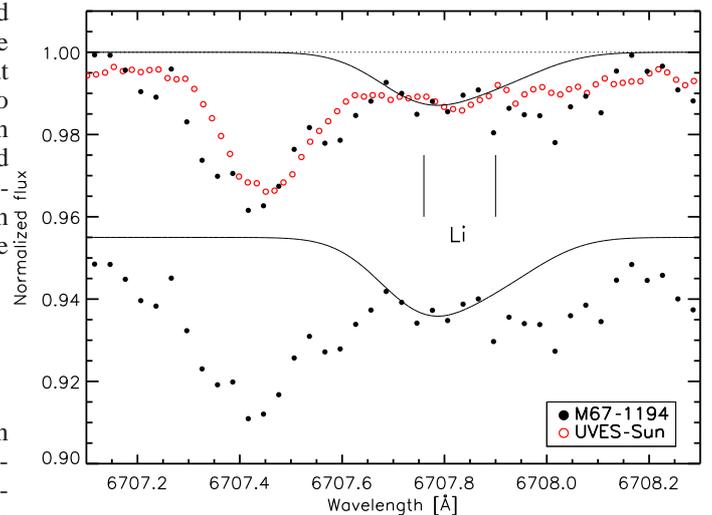}
  \caption{Upper part: The spectrum of M67-1194 (black bullets) around 
    Li\,{\sc i} 6707\,\AA\ with a local continuum compared to the
    FLAMES-UVES spectrum representing the Sun (red circles). 
    The best theoretical fit to the spectrum is shown as a solid line.
    Lower part: A global continuum placement for M67-1194 together
    with the best theoretical fit (both offset vertically by 0.045 units).}
  \label{fig:li_zoom}
\end{figure}

Pasquini et al. (2008) estimated the lithium content of ten solar-type stars
in M67 including M67-1194 from FLAMES-MEDUSA data. The authors derived
lithium abundances similar to that of the Sun from the Li doublet at
6707.8\,\AA, using spectra with a resolution of $R$\,$\sim$\,17,000
and a S/N ratio of 80--110\,pixel$^{-1}$.
The weakness of the Li\,{\sc i} 6707.8\,\AA\ doublet in M67-1194 and the 
noise level of the spectra make it difficult to determine an accurate
Li abundance for the twin. A major problem is the location of the local
continuum around the line. With our preferred positioning of the
continuum (cf. Fig.~\ref{fig:li_zoom}, lower part) we find a stronger 
6707\,\AA\
line than the solar one, and a Li abundance of log $\varepsilon$ = 1.26, 
corresponding to [Li/H] = +0.2 dex. This continuum positioning presumes that 
the wide depression around 6707.8\,\AA\ is a real spectral feature, and not 
due to some instrumental effect. If the latter is assumed, a more local 
continuum drawing is appropriate
(see upper part of Fig.~\ref{fig:li_zoom}).
This would lead to a Li abundance close to solar, [Li/H]\,=\,0.0,
i.e. a value closer but still greater than that obtained by Pasquini et al. 
(2008), [Li/H]\,$<$\,--0.2. We regard the Li\,{\sc i} doublet line 
identification in 
M67-1194 to be a probable but not a definitive one. It can therefore not be 
excluded that the true abundance is significantly below solar. 
Probably, however, the Li abundance of M67-1194 is solar or larger, 
but not exceeding the solar value by more than a factor of 2.5.

\section{The age of M67}
\label{sec:age}
Analysing a target very similar to the Sun, we profit from
a number of advantages.
The spectroscopic stellar parameters are more accurately determined
than is generally achieved.
This is again a result of the strictly differential approach relative
to the Sun, which is expected to promote
cancellation of the modelling errors that usually dominate
these types of analyses. The absolute accuracy
in the values obtained is directly based on the accurate
knowledge of the solar parameters.

This is then also the virtue of our age determinations. The
stellar-evolutionary tracks are admittedly uncertain, due to e.g.\ diffusion,
mixing and overshooting, the choice of mixing length parameter ($\alpha$) and 
He mass fraction (Y), as well as the absolute values of the solar CNO
abundances. However, $\alpha$ and Y are set to fit the solar radius
and age. Assuming that $\alpha$ and Y are close to solar
also for the solar twin, we may then use the tracks calibrated in this
way to obtain the age of the twin and thus of M67.

\begin{figure}
  \hspace{-0.6cm}
  \centering
   \includegraphics[width=8.0cm]{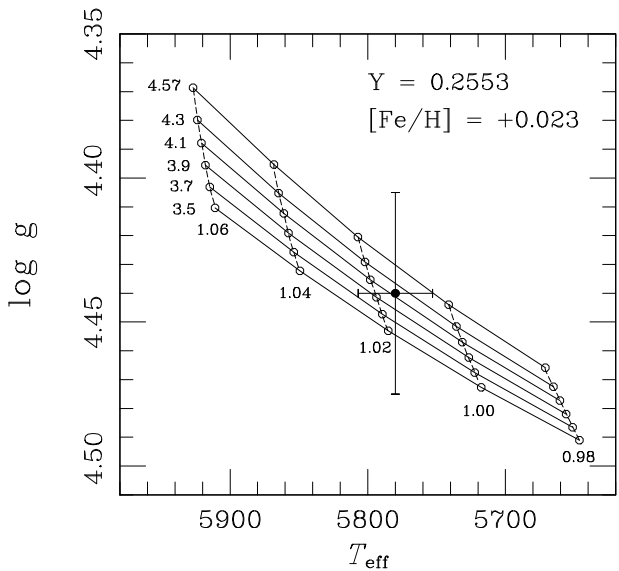}
  \caption{Plot of the 3.5 to 4.57\,Gyr segments of the post-main-sequence 
   tracks (dashed) that
   were computed for masses from 0.98 to 1.06\,${{\cal M}_\odot}$ 
   (indicated below each track) and the
   indicated values of $Y$ and [Fe/H].  Open circles represent the
   $T_{\rm eff} and \log\,g$ values at the ages noted just to the right of the
   $y$-axis; and hence the solid curves are isochrones for those ages.  The
   filled circle and attached error bars indicate the location of the solar
   twin, M$\,$67-1194.}
  \label{fig:iso}
\end{figure}

Using the current version of the Victoria stellar evolutionary code, we 
have computed several grids of tracks in order to evaluate the dependence of 
the predicted age on $T_{\rm eff}$, $\log\,g$, $Y$ and [Fe/H].  The main
improvements that have been made to this code since the VandenBerg et 
al.~(2007)
study of M$\,$67 are the following (1) The gravitational settling of helium 
and turbulent mixing below envelope convection zones are now treated, using 
methods very similar to those employed by Proffitt \&
Michaud (1991)\footnotemark, (2) the latest rates
for H-burning reactions (see Weiss 2008, Marta et al.~2008) and the improved
conductive opacities given by Cassisi et al.~(2007) have been adopted and (3)
the assumed solar abundances are those reported by Asplund et al.~(2009).  To
reproduce the properties of the Sun at its present age (4.57 Gyr; Bahcall et
al.~2005), it is necessary to adopt an initial helium content corresponding to
$Y = 0.2553$, along with a value of $\alpha_{\rm MLT} = 2.007$ for the usual
mixing-length parameter. 
\footnotetext{
At the solar age, settling reduces the surface 
helium abundance to $Y = 0.2384$, which is slightly less than the value 
inferred from
helioseismology, but similar to the values predicted by other Standard Solar
Models, see the Bahcall et al.~paper.  Following Proffitt \& Michaud, the
free parameter in the adopted treatment of turbulent mixing was set so that our
model for the Sun predicts the observed solar Li abundance.}
The evolutionary tracks in each grid, for masses in the range $0.98
\leq {\cal M}/{\cal M)_\odot} \leq 1.06$, were evolved from fully-convective
structures high on the Hayashi line to the solar age.  Interpolations 
between the tracks yielded values of $T_{\rm eff}$ and $\log\,g$ at ages 
ranging from 3.5 to 4.57 Gyr, which encompasses most of the ages that have 
been determined for M$\,$67 in the past decade (see below).  An example of 
one of the computed grids on the $(T_{\rm eff},\,\log\,g)$-plane is shown in 
Fig.~\ref{fig:iso}.  

From cubic spline interpolations in these results, the inferred age of the 
solar twin is 4.21 Gyr and its mass is estimated to be 1.013 
${{\cal M}_\odot}$. At $T_{\rm eff} =
5780$~K, one can readily determine that the gravity uncertainty implies a 
range in age from $\approx 2.6$
to 5.6 Gyr.  The age range spanned by the $T_{\rm eff}$ error bar is much
smaller, 3.8--4.6 Gyr.  Similar analyses of the model grids which were computed
for slightly different helium and metal abundances yield the results that are
listed in Table~\ref{tab:errorage}.  The measured uncertainty in [Fe/H] 
($\pm 0.015$) implies an
age uncertainty of approximately $\pm 0.2$ Gyr, whereas the evolutionary 
calculations predict about a $\pm 0.3$ Gyr change in age if the assumed value 
of $Y$ is altered by $\pm 0.01$. Mel{\'e}ndez et al. (2009) propose  that the 
Sun could 
have a reduced  fraction of refractories in its present relatively
thin convection zone, as compared with the composition of its  
interior. The effects of such a chemical inhomogeneity not only for
the Sun but also for M67-1194 on its age determination would cancel,  
but it is also possible that the twin, in spite of its solar abundance profile 
in the atmosphere, does not have a corresponding inhomogeneity. In that  
case, we find (using the results given in the bottom row of 
Table~\ref{tab:errorage}) a systematic error in the age determination, leading 
to an underestimate of the age by about 0.35\,Gyr.
Table~\ref{tab:age} provides a convenient summary of the errors in
the fundamental parameters of M$\,$67-1194 and their effects
on its inferred mass and age.

Our best estimate of the age of M$\,$67-1194 (4.2\,$\pm$\,1.6 Gyr) agrees very 
well with
with determinations based on alternative methods.  Most fits of isochrones to
the M$\,$67 color-magnitude diagram, whether using Garching (see Magic et
al.~2010), Montreal (Michaud et al.~2004), Padova (Yadav et al.~2008), Teramo
(Bellini et al.~2010), Victoria (VandenBerg et al.~2007), or Yale-Yonsei
(Yadav et al.) isochrones, favour an age of $4.0\pm 0.3$ Gyr.  Moreover,
Bellini et al.~have recently found that the age obtained from the cluster white
dwarfs is fully consistent with the aforementioned turnoff age.  However,
higher ages cannot yet be completely excluded (see examples provided by Yadav
et al.~and Magic et al.).  We note that the solar age is within
the $2\sigma$ error bar of M67's age.

\begin{table}[h]
  \caption{The chemical-composition dependence of the inferred mass and age of 
    M67-1194.}
  \centering
  \begin{tabular}[h]{lcccc}
[Fe/H]   &  $z$    &     Y    &   ${\cal M}$ [${\cal M}_{\odot}$] &  Age [Gyr]\\
     \hline \hline
     0.0   &   0.01323 &   0.2553  &   1.002  &  4.43 \\
     0.023 &   0.01413 &   0.2450  &   1.039  &  3.87 \\
     0.023 &   0.01393 &   0.2553  &   1.014  &  4.17 \\
     0.023 &   0.01375 &   0.2650  &   0.991  &  4.44 \\
     0.023 &   0.01656\tablefootmark{a} &   0.2553  &   1.028  &  3.50 \\
     0.046 &   0.01468 &   0.2553  &   1.027  &  3.91 \\
     \hline
   \end{tabular} 
   \tablefoot{\tablefoottext{a}{A mixture with enhanced CNO
       abundances; specifically, [C/Fe] = [N/Fe] = [O/Fe] = 0.10.}}
   \label{tab:errorage}
\end{table}

\begin{table}[h]
  \caption{Errors ($\epsilon$) in mass and age estimates for M67-1194 due to
    uncertainties in fundamental stellar parameters.}
  \centering
  \begin{tabular}[h]{clcc}
     Fund. param. & Value & $\epsilon({\cal M}/{\cal M}_{\odot})$ &  $\epsilon(Age)$ in Gyr \\
     \hline \hline
      \teff & 5780\,$\pm$\,27\,K      &  $\pm$\,0.01  &  $\mp$\,0.4 \\
     $\log g$ & 4.44\,$\pm$\,0.035  &  $\pm$\,0.01  &  $\mp$\,1.5  \\
     $\rm [Fe/H]$ & 0.023\,$\pm$\,0.015  & $\pm$\,0.01   & $\mp$\,0.2 \\
     Y & 0.2553\,$\pm$\,0.013      & $\mp$\,0.01   & $\pm$\,0.3  \\
     \hline
  \end{tabular}
  \label{tab:age}
\end{table}

\section{Discussion}
\label{sec:disc}
Our main finding, illustrated in Fig.~\ref{fig:diff}, is that M67-1194 is
indeed more similar to the Sun than most of the nearby solar twins in the field. This is interesting in view of the possible explanations
for the systematic departures of the field twin from the Sun in abundances with
condensation temperature, found and discussed by
Mel\'endez et al. (2009, see also Gustafsson et al.\ 2010). Basically, three 
possible explanations were discussed:
(1) The solar proto-planetary gas disk was sufficiently long-lived,
after the gas had been depleted in refractories by the formation of
planetesimals and planets, for the solar convection zone to become
so thin that the accreted gas could give visible traces in the solar
spectrum.
The field twins, however, in general accreted their gas disks
much earlier. (2) The solar outer convection zone never reached
deep enough to encompass most of the solar mass, and was therefore
much more easily polluted by the rarified infalling disk. This
property of the Sun, 
presumably reflecting the initial conditions of
the cloud to later form the Sun and the Solar System or the details of 
the more or less episodic accretion history of the Sun (cf. Baraffe \& 
Chabrier 2010), was not
usual among solar-type stars and is thus not shared by
most twins. (3) The proto-solar cloud was
cleansed by radiation fields from massive stars, pushing dust grains
out of the cloud. These effects must then have been
greater for the solar cloud than for most of the twins.

If the tendency found here for our M67 twin is
representative for the cluster, it speaks against the first
explanation above. We see no reason why the gas disks around stars in a
cluster should have longer life-times than those in the field.
It might be that the initial conditions of star formation are
different in a cluster than in the field, and that this might couple
to the evolution of the stellar convection zones, but this is only
speculation. 
Maybe, a more episodic
accretion history, gradually building stars from quite small entities,  
is more characteristic of cluster environment. In any case,
there is
independent evidence from radioisotopic abundances in the solar
system that the Sun was born in a cluster (Looney et al. 2006, see also
Portegies Zwart, 2009) containing relatively close supernovae or massive
AGB stars (Trigo-Rodr{\'i}guez et al. 2009).

The similarity of the age and overall composition of the Sun with the
corresponding data of M67,
and in particular the agreement of the detailed chemical composition
of the Sun with that of M67-1194, could suggest that the Sun has
formed in this very
cluster. According to the numerical simulations by Hurley et al.
(2005) the cluster has lost more than 80\% of its stars by tidal
interaction
with  the Galaxy, in particular when
passing the Galactic plane, and the Sun might be one of those. We note
that the orbit of the
cluster encloses, within its apocentre and pericentre, the solar orbit.
However, the cluster has an orbit extending to much higher Galactic
latitudes, presently it is close to its vertical apex at $z$ = 0.41\,kpc
(Davenport \& Sandquist 2010), while the Sun does
not reach beyond $z$ = 80\,pc  (Innanen, Patrick \& Duley 1978).
Thus, in order for this hypothesis of an M67
origin of the Sun to be valid, it must have been dispersed from the
cluster into an orbit precisely in
the plane of the Galactic disk, which seems improbable.

\section{Conclusions}
\label{sec:concl}

We draw the following conclusions from the analysis of M67-1194
presented here:

\begin{itemize}

\item
Within the framework of a differential analysis, the stellar
parameters (\teff, $\log g$, mass and age) of
M67-1194 are found to be compatible with the solar values,
while [Fe/H] is found to be slightly super-solar.
M67-1194 is thus the first solar twin known to belong to a
stellar association.
\item
The chemical abundance pattern of M67-1194 closely resembles the
solar one, in contrast to most known solar twins. This suggests
similarities in the formation of M67-1194 and the Sun. A common
origin of both stars as members of M67 is conceivable, albeit not
likely, considering their different galactic orbits. If the chemical
abundance pattern reflects environmental effects, then the Sun was
likely born in a cluster similar to M67.
\item
Of the three scenarios for the peculiar abundance pattern of the Sun
relative to field solar twins presented in Mel\'endez et al. (2009), a
particularly long-lived gas disk is the least probable due to the
cluster membership of M67-1194. In the standard picture of stellar
evolution with an early fully-convective phase, dust cleansing by
nearby luminous stars seems to have affected both the Sun and
M67-1194. Another possibility is that the initial boundary conditions
or accretion history of the early Sun were similar to those in the 
young cluster, and that these led to a shallow convection zone in the early 
evolution of these stars. More stars in M67
must be investigated to give further support to such explanations.
\end{itemize}

\noindent
Methodologically, we consider our method of age determination, based on
spectroscopy
of solar-like stars, to be an interesting alternative to
other determinations of cluster ages. Its main advantages is that the
uncertainties in
cluster distance and reddening are not important.
The age errors, in particular due to errors in the surface gravity, could be
further reduced by higher S/N, a wider spectral range and observations
of more stars. The uncertainty due to the unknown He abundance could be
reduced by accurate studies of the horizontal branch or of visual binaries.
In fact, the method can be extended to age estimates
for  other clusters, including globular ones. The basic idea is to make
a systematic spectroscopic comparison with well known standard
stars with very similar parameters in the nearby field or in other
clusters. Even if the absolute ages of these standards are not as
well established, the method offers an accurate age
ranking, revealing possible age differences between different
populations of clusters.

In general, the highly differential method used here has great
possibilities in other situations when detailed comparisons are to be 
carried out e.g. between Thick  and Thin Galactic-disk stars, or different 
types of Halo stars. Then, a close adherence to a pairwise selection
of stars with similar effective temperatures and metallicities
will make it possible to keep the systematic modelling
errors at a low level.

\acknowledgements
{We thank the anonymous referee whose constructive criticism considerably 
improved the paper. A{\"O} acknowledges support by V{\"a}rmlands nation 
(Uddeholms research scholarship). AJK acknowledges support by the Swedish 
Research Council and the Swedish National Space Board.} 

\thebibliography{}{
\bibliographystyle{astron}
\bibliography{mnemonic,ref_C}
\bibitem[]{}
Asplund, M., Grevesse, N., Sauval, A.~J., \& Scott, P.~2009, ARA\&A, 47, 481
\bibitem[]{}
Bahcall, J.~N., Basu, S., Pinsonneault, M., \& Serenelli, A.~M.~2005, ApJ,
  618, 1049
\bibitem[]{}
Baraffe, I. \& Chabrier, G. 2010, arXiv:~1008.4288v1
\bibitem[]{}
Baumann, P., Ram{\'i}rez, I., Mel{\'e}ndez, J., Asplund, M., 
Lind, K. 2010, arXiv:~1008.0575 
\bibitem[]{}
Biazzo K., Pasquini, L., Bonifacio, P., Randich, S. \& Bedin, L. R. 2009,
 MmSAI, 80, 125
\bibitem[]{}
 Bellini, A. et al. 2010, A\&A, 513, A50
\bibitem[]{}
Casagrande, L., Ram{\'i}rez, I., Mel{\'e}ndez, J., Bessell, M.,
Asplund, M. 2010, A\&A, 512, 54 
\bibitem[]{}
Cassisi, S., Potekhin, A., Pietrinferni, A., Catelan, M., \& Salaris, M. 2007,
  ApJ, 661, 1094
\bibitem[]{}
Cayrel de Strobel, G., Knowles, N., Hernandez, G., \& Bentolila, C. 1981,
 A\&A, 94, 1
\bibitem[]{}
Cayrel de Strobel, G. 1996, A\&ARv, 7, 243
\bibitem[]{}
Cox, A. N. (ed.) 2000, Allen's Astrophysical Quantities, Fourth Ed., 
 Springer-Verlag New York, Inc.
\bibitem[]{}
Davenport, J. R. A. \& Sandquist, E. L. 2010, ApJ, 711, 559
\bibitem[]{}
Fuhrmann, K., Pfeiffer, M., Frank, C., Reetz, J., \& Gehren, T. 1997, A\&A,
 323, 909
\bibitem[]{}
Gray, D. F. 1994, PASP, 106, 1248
\bibitem[]{}
Grupp, F. 2004, A\&A, 420, 289
\bibitem[]{}
Gustafsson, B. 1998, SSrv, 85, 419
\bibitem[]{}
Gustafsson, B., Edvardsson, B., Eriksson, K., J\o rgensen, U. G., Nordlund,
 \AA., \& Plez, B. 2008, A\&A, 486, 951
\bibitem[]{}
Gustafsson, B., Mel{\'e}ndez, J., Asplund, M., Yong, D. 2010, Ap\&SS, 328, 185
\bibitem[]{}
Hardorp, J. 1978, A\&A, 63, 383
\bibitem[]{}
Hobbs, L. M. \& Thorburn, J. A. 1991, AJ, 102, 1070
\bibitem[]{}
Holmberg, J., Flynn, C., \& Portinari, L. 2006, MNRAS, 367, 449
\bibitem[]{}
Hurley J. R., Pols, O. R., Aarseth, S. J., \& Tout, C. A. 2005, MNRAS, 363, 293
\bibitem[]{}
Innanen, K. A., Patrick, A. T., Duley, W. W. 1978, Ap\&SS, 57, 511
\bibitem[]{}
Korn, A. J., Shi, J., \& Gehren, T. 2003, A\&A, 407, 691
\bibitem[]{}
Kurucz, R. L., Furenlid, I., Brault, J., \& Testerman, L. 1984, Solar
 Flux Atlas from 296 to 1300 nm, Kitt Peak National Solar Observatory
\bibitem[]{}
Lodders, K. 2003, ApJ, 591, 1220
\bibitem[]{}
Looney, L. W., Tobin, J. J., \& Fields, B. D. 2006, 652, 1755
\bibitem[]{}
Magic, Z., Serenelli, A.~M., Weiss, A., \& Chaboyer, B.~2010, ApJ, 718, 1378
\bibitem[]{}
Marta, M., et al. 2008, ~Phys.~Rev.~C, 78, 022802
\bibitem[]{}
Mel\'endez, J., Dodds-Eden, K., \& Robles, J. 2006, ApJ, 641, 133
\bibitem[]{}
Mel\'endez, J. \& Ram\'irez, I. 2007, ApJ, 669, 89
\bibitem[]{}
Mel\'endez, J., Asplund, M., Gustafsson, B., \& Yong, D. 2009, ApJ, 704, L66
\bibitem[]{}
Michaud, G., Richard, O., Richer, J., \& VandenBerg, D.~A.~2004, ApJ, 606, 452
\bibitem[]{}
Pace, G., Pasquini, L., \& Fran\c{c}ois, P. 2008, A\&A, 489, 403
\bibitem[]{}
Pasquini, L., Randich, S., \& Pallavicini, R. 1997, A\&A, 325, 535
\bibitem[]{}
Pasquini, L. Biazzo, K., Bonifacio, P., Randich, S., \& Bedin, L. R. 2008,
 A\&A, 489, 677
\bibitem[]{}
Piskunov, N. E. \& Valenti J. A.  2002, A\&A 385, 1095
\bibitem[]{}
Portegies Zwart, S. F. 2009, ApJ, 696, 13
\bibitem[]{}
Porto de Mello, G. F., da Silva, R., \& da Silva, L. 2000, ASPC, 213, 73
\bibitem[]{}
Porto de Mello, G. F. \& da Silva, L. 1997, ApJ, 482, 89
\bibitem[]{}
Ram\'irez, I., Mel\'endez J., \& Asplund, M. 2009, A\&A, 508, 17
\bibitem[]{}
Proffitt, C.~R., \& Michaud, G.~1991, ApJ, 371, 584
\bibitem[]{}
Randich, S., Sestito, P., Primas, F., Pallavicini, R., \& Pasquini, L. 2006,
 A\&A, 450, 557
\bibitem[]{}
Reetz, J. K. 1991, Diploma Thesis, Universit\"at M\"unchen
\bibitem[]{}
Takeda, Y., Kawanomoto, S., Honda, S., Ando, H., \& Sakurai, T. 2007, A\&A,
468, 663
\bibitem[]{}
Taylor, B. J. 2007, AJ, 133, 370
\bibitem[]{}
Tautvai$\check{\rm s}$iene, G., Edvardsson, B., Tuominen, I., \& Ilyin, I.
 2000, A\&A, 360, 499
\bibitem[]{}
Trigo-Rodr\'iguez, J. M., Garc\'ia-Hern\'andez, D. A., Lugaro, M.,
  Karakas, A. I.,van Raai, M., Garc\'ia Lario, P., \& Manchado, A. 2009,
  M\&PS, 44, 627
\bibitem[]{}
VandenBerg, D. A., Gustafsson, B., Edvardsson B., Eriksson, K. \&
 Ferguson, J. 2007, ApJ, 666, 105
\bibitem[]{}
Weiss, A.~2008, Physica Scripta, T133, 014025
\bibitem[]{}
Yadav, R. K. S, Bedin, L. R., Piotto, G., Anderson, J., Cassisi, S., Villanova,
S., Platais, I., Pasquini, L., Momany, Y. \& Sagar, R. 2008, A\&A, 484, 609
\bibitem[]{}
Yong D., Carney, B. W., \& Teixera de Almeida, M. L. 2005, AJ, 130, 597
}

\newpage

\begin{appendix}
\section{Chemical elements}

\begin{table}[h]
  \caption{Fe\,{\sc i} and Fe\,{\sc ii} lines used in the analysis.}
  \centering
  \begin{tabular}[h]{cc|cc}
    \textbf{Fe\,{\sc i}} & \textbf{[Fe/H]} & \textbf{Fe\,{\sc ii}}&
    \textbf{[Fe/H]} \\
   \hline \hline
      5166.281 &   0.02   & 5197.577 &   0.02  \\
      5198.711 &   0.02   & 5234.625 &   0.03  \\
      5216.274 &   0.02   & 5264.812 &   0.04  \\
      5217.389 &   0.04   & 5284.109 &   0.01  \\
      5242.491 &   0.05   & 5325.553 &   0.00  \\
      5247.049 &   0.05   & 5414.073 &   0.04  \\
      5225.535 &   0.02   & 5425.257 &   0.02  \\
      5250.208 &   0.03   & 6247.557 &   0.03  \\
      5295.299 &   0.01   & 6456.383 &   0.03  \\
      5367.466 &   0.00   &          &         \\
      5379.574 &   0.02   &          &         \\
      5522.446 &   0.04   &          &         \\
      5638.262 &  -0.01   &          &         \\
      5662.516 &   0.03   &          &         \\
      5679.023 &   0.01   &          &         \\
      5701.514 &   0.02   &          &         \\
      6065.482 &   0.02   &          &         \\
      6151.617 &   0.01   &          &         \\
      6200.313 &   0.01   &          &         \\
      6213.429 &   0.04   &          &         \\
      6229.225 &   0.02   &          &         \\
      6252.554 &   0.04   &          &         \\
      6421.349 &   0.02   &          &         \\
      6498.937 &   0.01   &          &         \\
      \hline
  \end{tabular}
\end{table}

\begin{figure}
  \hspace{-0.6cm}
  \centering
   \includegraphics[width=10.0cm,angle=90]{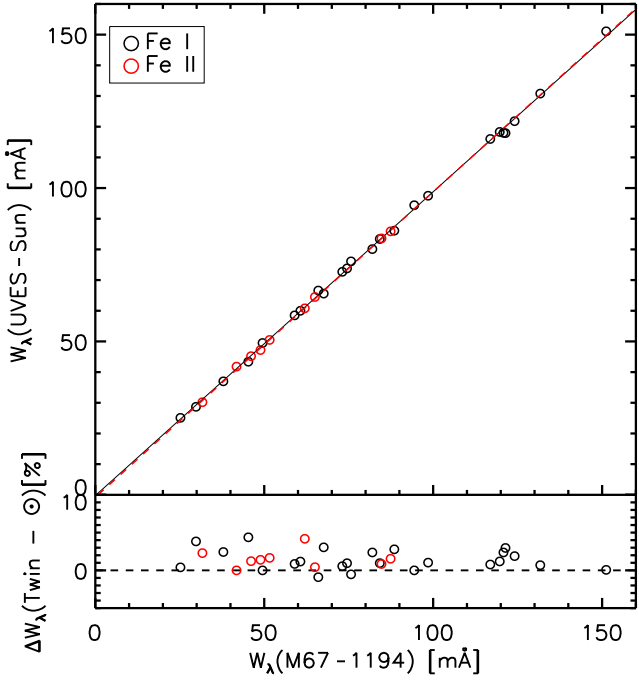}
  \caption{Upper panel: Comparison of equivalent widths of M67-1194 and the 
    FLAMES/UVES sky spectrum representing the Sun. 1D polynomial fits for 
    both Fe\,{\sc i} and Fe\,{\sc ii} are shown
    as solid and dashed lines, respectively.
    Lower panel: The percentage equivalent-width difference of Fe\,{\sc i} and 
    Fe\,{\sc ii}
    lines between M67-1194 and the FLAMES/UVES sky spectrum representing the 
    Sun. Note that the abundances are not derived from equivalent 
    widths, but from fits of synthetic spectra.}
\end{figure}

\begin{table}[h!]
  \caption{Chemical elements (except Fe) and spectral lines in
    the analysis.}
  \centering
  \begin{tabular}[h]{cr|cr}
     \textbf{Element} & \textbf{$\Delta$(Sun -- M67-1194)} & \textbf{Element}& \textbf{$\Delta$(Sun -- M67-1194)}\\
     \hline \hline
    \textbf{C\,{\sc i} \& C\boldmath{$_2$}\unboldmath} &   & \textbf{V\,{\sc i}}    & \\
     5380.337       &      0.03   &  5670.853       &     0.02    \\
     5135.570       &      0.01  &  6039.722       &     0.05    \\
     5141.210       &      0.00   &  6081.441       &     0.01    \\
     5147.694       &      0.015  &  6090.214       &   --0.01    \\
     \textbf{O\,{\sc i}}   &             &  6199.197       &   --0.03    \\
     6158.190       &      --0.07   &  \textbf{Cr\,{\sc i}}  &             \\
     \textbf{Na\,{\sc i}}  &             &  5238.964       &     0.00    \\
     6154.225       &      0.03   &  5247.566       &   --0.02    \\
     6160.747       &      0.03   &  5272.007       &     0.05    \\
     \textbf{Mg\,{\sc i}}  &         &  5287.200       &     0.04    \\
     5711.088       &    --0.01   &  5296.691       &   --0.03    \\
     6318.710\tablefootmark{b} &      0.00   &  5300.744       &     0.00    \\
     6319.240       &    --0.04   &  \textbf{Mn\,{\sc i}}  &         \\
     \textbf{Al\,{\sc i}}  &             &  6013.513       &     0.00    \\
     6696.018       &     0.00    &  6016.673       &   --0.01    \\
     6698.667       &     0.02    &  6021.819       &     0.01    \\
     \textbf{Si\,{\sc i}}  &         &  \textbf{Co\,{\sc i}}  &         \\
     5488.983       &     0.02    &  5247.911       &     0.02    \\
     5665.554       &   --0.02    &  5301.039       &   --0.015   \\
     5690.425       &     0.08    &  5342.695       &     0.02    \\
     5701.104       &   --0.02    &  5530.774       &   --0.01    \\
     6125.021       &   --0.02    &  5647.230       &     0.02    \\
     6145.015       &     0.05    &  6189.000       &     0.04    \\
     6243.823       &   --0.015   &  \textbf{Ni\,{\sc i}}  &             \\
     6244.476       &   --0.03    &  5589.358       &     0.01    \\
     6721.848       &     0.01    &  5643.078       &   --0.01    \\
     6741.630       &     0.005   &  6086.282       &     0.02    \\
     \textbf{S\,{\sc i}}   &         &  6108.116       &     0.00    \\
     6046.000       &     0.005   &  6130.135       &     0.005   \\
     6052.656       &     0.02    &  6204.604       &     0.00    \\
     6743.540       &     0.01    &  6223.984       &     0.03    \\
     6757.153       &   --0.02    &  6767.772       &   --0.02    \\
     \textbf{Ca\,{\sc i}}  &         &  6772.315       &     0.02    \\
     5512.980       &   --0.02    &  \textbf{Cu\,{\sc i}}  &             \\
     5590.114       &   --0.02    &  5218.197       &    0.00     \\
     5867.562       &     0.05    &  5220.066       &  --0.005    \\
     6166.439       &   --0.02    &  \textbf{Y\,{\sc ii}}  &             \\
     6169.042       &   --0.01    &  4854.867       &   --0.07   \\
     \textbf{Sc\,{\sc ii}} &             &  4883.685       &   --0.01   \\
     5657.896       &     0.00    &  4900.110       &   --0.04   \\
     5684.202       &   --0.02    &  5087.420       &     0.04   \\
     6245.637       &     0.02    &                 &             \\
     \textbf{Ti\,{\sc i}}  &         &                 &             \\
     5113.448       &     0.02    &                 &             \\
     5219.700       &   --0.01    &                 &             \\
     5490.150       &     0.00    &                 &             \\
     5866.452       &     0.01    &                 &             \\
     6126.217       &     0.00    &                 &             \\
     6258.104       &   --0.01    &                 &             \\
     6261.101       &     0.05    &                 &             \\
     \hline
   \end{tabular}
   \tablefoot{\tablefoottext{b}{For this line the KPNO atlas was used to represent the 
       Sun as the FLAMES-UVES spectrum suffered from telluric absorption at this
       wavelength.}}
 \end{table}

\end{appendix}

\end{document}